\documentclass[pre,reprint,nofootinbib,superscriptaddress]{revtex4-1}
 \pdfoutput=1
\usepackage[pdftex]{graphicx}
\usepackage{amssymb,amsfonts,amsmath}
\usepackage{color}
\newcommand{\tw}[1]{{\color{blue}[\hbox{\lower.5ex\hbox{\tiny TW}} #1]}}
\newcommand{\ee}[1]{{\color{green}[\hbox{\lower.5ex\hbox{\tiny EE}} #1]}}

\renewcommand{\d}{\delta}
\newcommand{\ep}{\epsilon}

\newcommand{\hatr}{\mathbf{\hat{r}}}

\newcommand{\hatt}{\mathbf{\hat{t}}}

\newcommand{\hatn}{\mathbf{\hat{n}}}

\newcommand{\beq}{\begin{equation}}
\newcommand{\eeq}{\end{equation}}

\newcommand{\pd}{\partial}

\newcommand{\mymat}[1]{\begin{pmatrix} #1 \end{pmatrix}}

\begin{document}
\title{Confined disclinations:\\
exterior vs material constraints in developable thin elastic sheets}
\author{Efi Efrati}
\email{efi.efrati@weizmann.ac.il}
\affiliation{Department of Physics of Complex Systems, Weizmann Institute of Science.
PO Box 26, Rehovot, 76100, Israel}
\affiliation{James Franck Institute, The University of Chicago, 929 E. 57th St, Chicago, IL 60637, USA}
\author{Luka Pocivavsek}
\affiliation{Department of Surgery, University of Pittsburgh, Pittsburgh, Pennsylvania}
\affiliation{James Franck Institute, The University of Chicago, 929 E. 57th St, Chicago, IL 60637, USA}
\author{Ruben Meza}
\affiliation{James Franck Institute, The University of Chicago, 929 E. 57th St, Chicago, IL 60637, USA}
\affiliation{Departamento de Fisica de la Universidad de Santiago de Chile, av. Ecuador 3493, Santiago, 9170124, Chile}
\author{Ka Yee C. Lee}
\affiliation{James Franck Institute, The University of Chicago, 929 E. 57th St, Chicago, IL 60637, USA}
\author{Thomas A. Witten}
\affiliation{James Franck Institute, The University of Chicago, 929 E. 57th St, Chicago, IL 60637, USA}

\begin{abstract}
We examine the shape change of a thin disk with an inserted wedge of material when it is pushed against a plane, using analytical, numerical and experimental methods. Such sheets occur in packaging, surgery and nanotechnology.  We approximate the sheet as having vanishing strain, so that it takes a conical form in which straight generators converge to a disclination singularity.  Then its shape is that which minimizes elastic bending energy alone.  Real sheets are expected to approach this limiting shape as their thickness approaches zero.  The planar constraint forces a sector of the sheet to buckle into the third dimension.  We find that the unbuckled sector is precisely semicircular, independent of the angle $\delta$ of the inserted wedge. We generalize the analysis to include conical as well as planar constraints and thereby establish a law of corresponding states for shallow cones of slope $\epsilon$ and thin wedges. In this regime the single parameter $\delta/\epsilon^2$ determines the shape.  We discuss the singular limit in which the cone becomes a plane. We discuss the unexpected slow convergence to the semicircular buckling seen experimentally.
\end{abstract}

\maketitle
 \section{Introduction} \label{Introduction}
Throughout history people have devised means of modifying two-dimensional materials, such as cloth or paper, to take on designed three-dimensional shapes \cite{wolff1996art}.  The art of tailoring consists of cutting and joining cloth to conform to the wearer's body.  To accomplish this the tailor forms disclinations by making a cut into the cloth and then inserting or removing a wedge-shaped sector, thus altering the intrinsic internal geometry of the material. The creation of the disclination introduces a point source of Gaussian curvature at the tip of the wedge. The amplitude or ``\textit{charge}'' of the point source gives, up to a sign, the wedge angle.
The point Gaussian charge forces the cloth to assume a non-flat three dimensional configuration.  The sheet may be further limited by the {\it exterior} constraint of impenetrability: the cloth cannot pass through itself or the wearer's body

Here we study the simplest ways that a sheet with a disclination defect can be altered by such an exterior constraint.  First we push a sheet containing an inserted wedge against an impenetrable plane.  We observe that the plane deforms the sheet in a novel and robust way. We then generalize to a sheet containing a point Gaussian charge of arbitrary sign confined by an impenetrable frictionless cone. Our results unify some of the results obtained for d-cones (confined flat sheets) \cite{CM98,CM05}, and e-cones (unconstrained disclinations in thin sheets) \cite{MBAG08,GMVM12}.  

These findings are part of a recent surge of interest in controlling the spatial configuration of a sheet by modifying its material geometry.
In the last decade, the art of controlling the spatial configuration of a sheet by modifying its internal construction has developed far beyond classical tailoring methodology. The discrete disclination charges of the tailor's art have been generalized to a continuum charge density profile of intrinsic Gaussian curvature.  In this way one can make materials that spontaneously form a range of designed shapes in response to temperature or solvents \cite{KES07,KHBSH12}.  The shapes arising from individual disclinations \cite{NS88,MBAG08} or multiple disclinations \cite{Wit07,GHKM13} are understood formally.  The shape selection is strongest in the asymptotic limit where the thickness of the material goes to zero relative to its other dimensions.  In this limit the strain in the surface approaches zero, and a disclination in an otherwise flat sheet approaches a conical shape in which every point is connected to the vertex by a straight, unbent line. These limiting configurations, termed conical defects, can be described mathematically in terms of elliptic functions.

A complementary means of shaping a sheet is to constrain its shape by external confinement.  The best studied example is the d-cone mentioned above.  As defined in Ref. \cite{CM05}, a d-cone is a flat sheet pushed into a conical container ({\it cf} \cite{CM98}).  Any surface fully conforming to the conical shape of the container would have a Gaussian charge, but the flat sheet has no such charge.  The sheet must accommodate this mismatch by buckling inward, away from the supporting cone, forming the d-cone structure.  
The angular size of the buckled region is selected by subtle matching conditions relating the solution in the buckled region and the conformation of the sheet in the supported region. In particular, even for an infinitesimally shallow cone, the ``takeoff lines" separating the free standing buckled region and the contacting region form a finite angle of about $139$ degrees.

\begin{figure}[ht]
\begin{center}
\includegraphics[width=8.6cm]{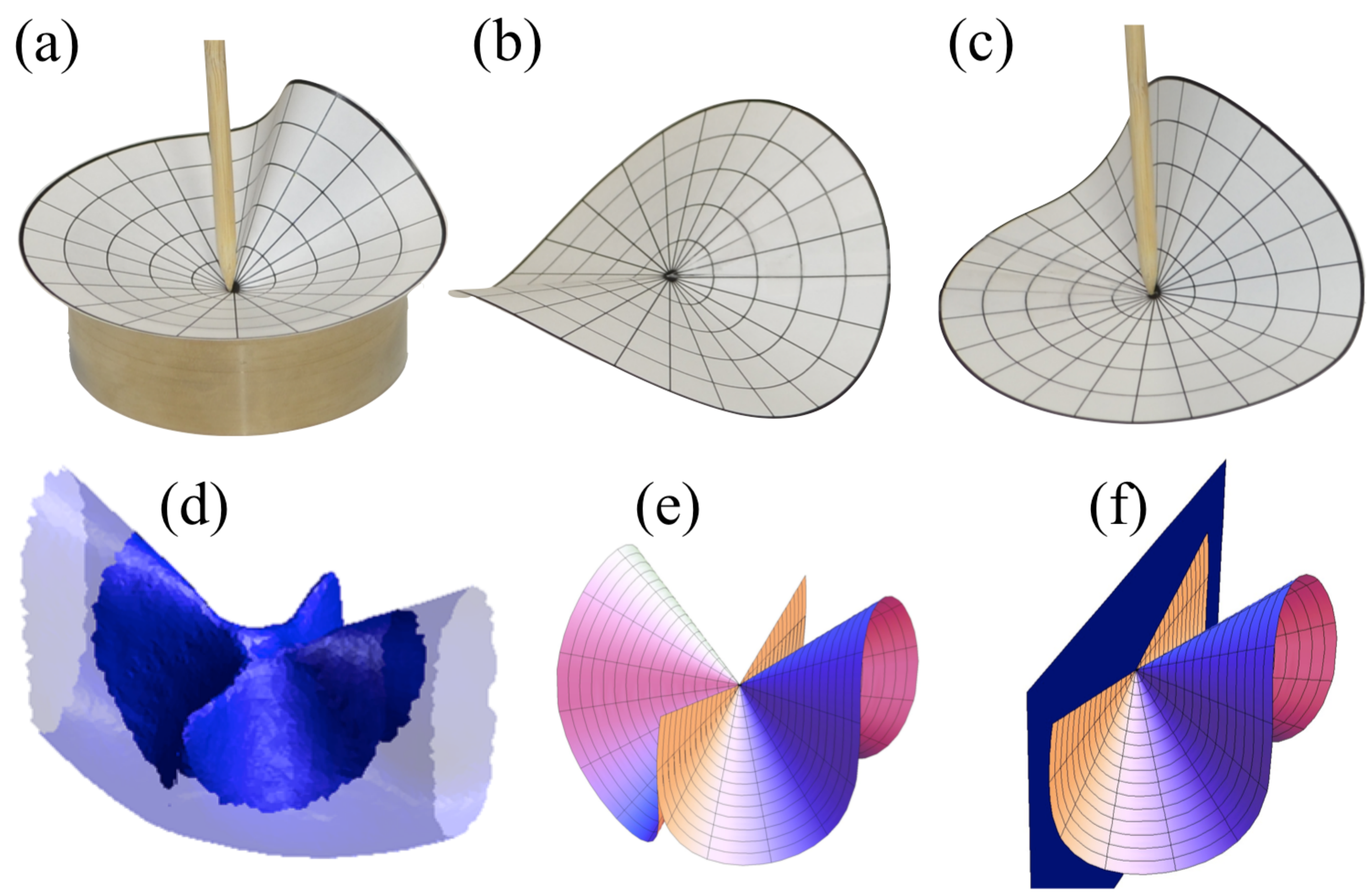}
\end{center}
\caption{\textit{Examples of conical defects}. (a-c):  photos of conical defects made of paper. (a) A d-cone. A flat disk is pushed by a point force at its center into a confining cone. (b) An unconstrained e-cone of Gaussian charge $-\pi/6$.
(c) A plane-supported conical defect. The e-cone appearing in (b) is pushed by a point force at its center onto a smooth supporting plane (not shown).  
(d) A reconstruction of a rubber tube subject to the longitudinal cutting and traverse rejoining as performed in the Heineke-Mikulicz surgery (see \cite{PELH13}).  The resulting geometry contains a $-2\pi$ Gaussian charge at the center of the joining line, and two $+\pi$ Gaussian charges at its ends.
 Shaded dark (blue) region notes the collection of points on the rubber tube separated by one radius or less from the center of the joining line. For more details see \cite{PELH13}. (e) The numerically calculated shape of an unconstrained strain-free sheet with a single conical defect of Gaussian charge $-2\pi$ along with two weak lines for bending (corresponding to the joining suture line) gives a very good agreement with the observed geometry in (d). 
(f) Due to the symmetry of the problem it can also be understood by studying the behavior of a conical defect supported on a frictionless impenetrable plane. Plotted is the right portion of the solution appearing in (e) along with its symmetry plane. The resulting structure is equivalent to (c) except for the magnitude of the disclination.}
\label{fig:cones}
\end{figure}

The two distinct mechanisms necessitating buckling: \textit{external} confinement, as in the d-cone of Fig \ref{fig:cones}(a), and \textit{internal geometry} (in the form of the intrinsic Gaussian charge) as in the  unconstrained e-cone of Fig \ref{fig:cones}(b), can occur simultaneously. 
An example of such a system is depicted in Fig \ref{fig:cones}(e) showing a plane supported wedge disclination. Such systems arise naturally in certain surgical reconstructions, and 
they show distinct behavior due to the simultaneous action of both effects, as discussed below.

\section{Examples of supported conical defects} \label{Examples}

In tayloring, the inserted wedge in an e-cone defect is called a godet.  Godets are inserted in order to add volume and flair to the bottom of a skirt. 
 If the godet wedge is fastened to a stiff, flat material, the added volume is increased. The confinement by a plane, shown in Figure \ref{fig:cones}(c) illustrates the same principle,  the confined sheet subtends a larger volume than the unconfined sheet in Figure \ref{fig:cones}(b).  A similar effect also appears in sheets plastically deformed near their edges \cite{Sharon:2002kx}.  This effect is used implicitly whenever a flat sheet is cut and joined in order to increase subtended volume.  Similar shapes emerge when a wrapper is torn open \cite{Hamm:2008zr} or a notched graphene sheet is stretched \cite{Blees:2014ys}.

Many surgical reconstructions amount to cutting an otherwise flat thin tissue (e.g. skin, slender muscle layers, etc.) and stitching the cut to itself so as to alter the material geometry. In much the same manner described above, this results in Gaussian charges. 
Examples of such procedures include the Heinicke Mikulitz (HM) strictureplasty performed to alleviate a bowel narrowing \cite{PELH13}, and the Z-plasty, a plastic surgery aimed at rotating the direction of an existing wound \cite{McG57,Fur65}. In both cases the resulting 2D geometry is that of a Gaussian charge quadrupole. In the latter procedure the skin's natural tensile stress and finite thickness keep the structure planar, while in the former procedure the surgically treated bowel assumes an expanded
three dimensional shape.

The structure resulting from the Heinicke Mikultz procedure, shown in Figure \ref{fig:cones}(c), is formed through the interplay of the geometry induced by disclinations created, thin sheet elasticity and impenetrability of the tissue itself \cite{PELH13}. As discussed in \cite{PELH13}, the negative point Gaussian charge at the center of the suture-line dominates the resulting shape.
Thus in order to understand the shape of a tube after HM strictureplasty it suffices to consider the simplified problem of two flat disks with a radial cut in each, connected to each other along the radial cut and splayed open, as depicted in  Figure \ref{fig:cones}(b). The outcome is a mirror-symmetric buckled shape in which each of the two disks displays a free-standing  buckled portion flanked by two flat regions supported on the flat region of the opposite disk. We can therefore further reduce the problem and solve the elastic equilibrium configuration of only one of the radially cut disks supported on a frictionless plane, as in  Figure \ref{fig:cones}(c). By this process we are led to the simplified problem of a plane supported conical defect: A thin elastic disk with a negative Gaussian charge (whose charge is given by the angle between the two edges of the radial cut) supported on a frictionless plane, as shown in Figure \ref{fig:glued}; we next determine this shape mathematically.

\section{Plane supported conical defect} \label{Plane}

The equilibrium shape of a thin conical defect is that which minimizes its bending energy.  For such conical configurations the surface is described by straight generators that all meet at a single vertex, {\it e.g.} the radial lines in Figs \ref{fig:cones}(a--c).  We specify position on the surface using radial and azimuthal coordinates $r$ and $s$. The configuration of such surfaces is determined uniquely by the non-vanishing component of their normal curvature $\kappa_1(r, s)$, whose variation along the generators is given by $\kappa_1(s,r)=\kappa(s)/r$. Equilibrium solutions for $\kappa(s)$ are available in terms of elliptic functions \cite{CM05,MBAG08}. For unconstrained disclinations (e-cones) the solution for
$\kappa(s)$ is periodic and differentiable.  For e-cones the period-two shape (in which the curvature has four nodes) seen in Figure \ref{fig:cones}(b) was shown to be the only stable solution for defects that do not have self contact \cite{MBAG08, DBA08,GMVM12}. For the case of a confined flat sheet, continuity of the curvature across the takeoff line, separating the buckled and supported portions of the sheet, leads to a solution in which the curvature possesses only two nodes, as seen in Figure \ref{fig:cones}(b) \cite{CM05}. We now specialize to the case of Fig. \ref{fig:cones}(c): a plane-supported e-cone.

We consider a conical defect made by inserting a sector of angle $\d$ into a disk of unit radius supported on a frictionless plane. The origin is set at the center of the disk such that the circumference of the disk lies on the unit sphere \cite{MBAG08}. We parameterize the edge of the disk by the unit vector $\hatr(s)$. We use an arc-length parameter $s$ such that $\hatr' \equiv \pd_s \hatr=\hatt$, is the unit tangent to the curve. The normal to the conical surface is given by $\hatn=\hatr\times\hatt$. These triply orthogonal unit vectors satisfy the following relation \cite{Wit07}:
\beq
\mymat{\hatr'\\ \hatt' \\ \hatn'}\equiv \frac{d}{d s}\mymat{\hatr\\ \hatt \\ \hatn}=\mymat{0&1&0\\-1&0&\kappa \\ 0&-\kappa &0}\mymat{\hatr\\ \hatt \\ \hatn}.
\label{eq:mat}
\eeq
Note that this is {\em not} the Serret-Frenet frame of the corresponding curve. In particular $\kappa=\hatt'\cdot(\hatr\times\hatt)$ is {\em not} the curvature of the curve given by $\hatr$ but rather the non-vanishing principal normal curvature of the conical surface bounded by the curve $\hatr$. Thus the bending energy is simply proportional to $\int \kappa^2 ds$  \cite{MBAG08}.

\begin{figure}[ht]
\begin{center}
\includegraphics[width=8.6cm]{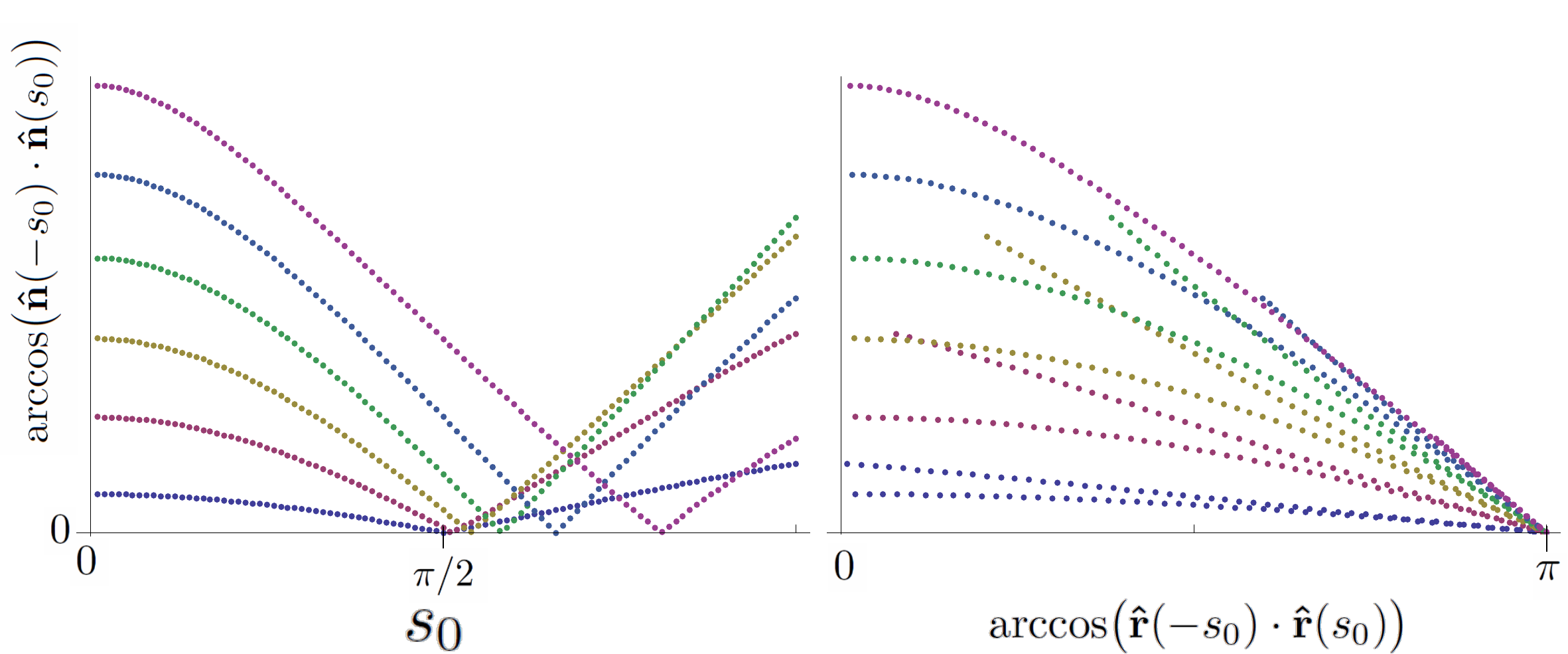}
\end{center}
\caption{\textit{Determining the takeoff angle:} Left: the angle $\theta_n$ between the normals as a function of positions $\pm s_0$ of vanishing curvature $\kappa$ for various  values of the parameter $c$, using Eq. \ref{eq:kgeneral}.  Each $c$ corresponds to a particular value of the wedge angle $\delta$, to be determined.  A valid takeoff point $s_0$ is one in which $\theta_n$ vanishes.  Right: the angle $\theta_n$ at the point of vanishing curvature plotted vs the angle $\theta_r$ between the displacement vectors  $\hat r(s_0)$ and $\hat r(-s_0)$.  It shows that at the valid takeoff point where $\theta_n=0$, $\theta_r = \pi$ for all values of $c$ plotted.  
}
\label{fig:takeoff-normals}
\end{figure}

\begin{figure*}[t]
\begin{center}
\includegraphics[width=16cm]{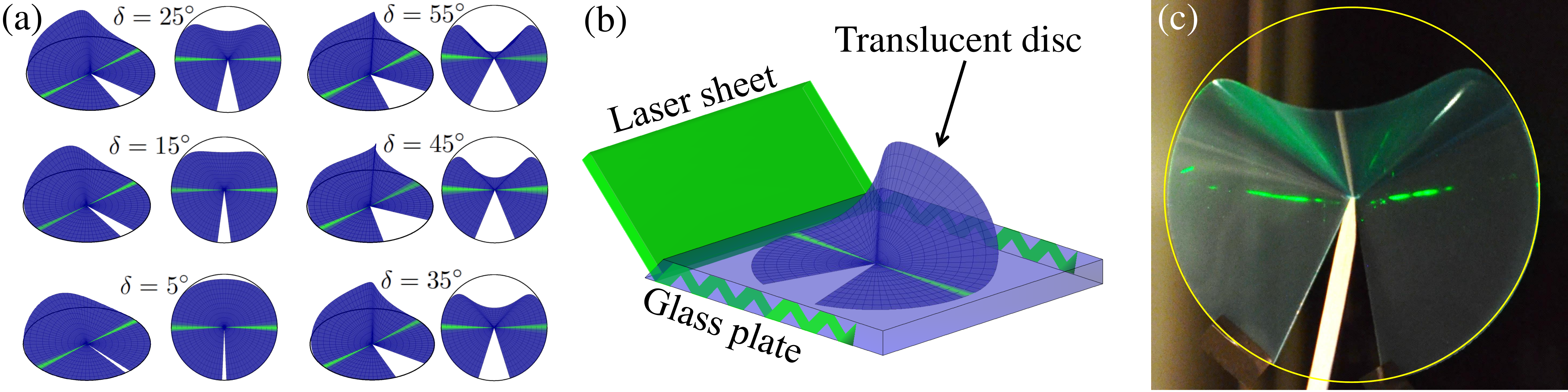}
\end{center}
\caption{\textit{Visualizing the takeoff angle in plane supported conical defects:} (a) Top and perspective view of simulated conical defects calculated as described in Sec. \ref{Plane} and colored by the normal force exerted on the thin sheet by the supporting plane; light (Green) shading implies greater normal force. The finite width of the bright (green) lines is due to the discretization and to non-zero compliance of the supporting plane. Note that varying the magnitude of the wedge angle
by an order of magnitude results in no discernible change in the takeoff angle, which remains at the value $180^\circ\pm 5^\circ$. (b) A schematic diagram of the total internal reflection (TIR) setup. A laser sheet enters into a glass slab at an angle such as to meet the glass air interface at an angle above the angle of TIR, reflecting entirely back into the glass. When a translucent plastic sheet touches the glass, light can travel into the plastic sheet and is then scattered in all directions. The amount of the light that enters the plastic sheet is proportional to the true area of contact and thus proportional to the local normal force between the plastic sheet and the glass plate \cite{Rubinstein:2004fk}. (c) Top view of a notched disk in a translucent thin plastic sheet imaged through TIR. 
Diameter was 300 mm ; thickness was 150 $\mu$m.
The white rod entering from the bottom serves to force down the plastic sheet at its vertex.
Yellow circle is shown to guide the eye.  Bright (green) regions correspond to the takeoff lines. 
The takeoff angle is somewhat smaller than the predicted $180$ degrees as discussed in the text.
}
\label{fig:glued}
\end{figure*}

With the aid of these vectors we may formulate the following elastic energy minimization problem in the free standing portion of the sheet:
\begin{eqnarray}
\label{eq:lagrangian}
E[\vec r, \vec t\,] =&&\mathop{\int_{-s_{0}}^{s_{0}}} ~ds~ \bigl( (\vec t\,'\cdot(\vec r\times\vec t\,))^2\\
&&+\lambda_r(\vec r\cdot\vec r - 1) + \lambda_t(\vec t\cdot\vec t - 1) + 2 \vec\eta\cdot (\vec r\,'-\vec t\,)\bigr).\nonumber
\end{eqnarray}

The first integrand is simply the normal curvature squared and accounts for the bending energy of the sheet. The two scalar Lagrange multiplier functions  $\lambda_t(s),\lambda_r(s)$ enforce unit length for the vectors $\hatr$ and $\hatt$ and can be related to the radial and azimuthal stresses. The Lagrange multiplier vector function $\boldsymbol{\vec \eta}(s)$ enforces the relation between $\hatt$ and $\hatr$. The use of unrestricted vector variables $\vec r$, $\vec t$ allows to easily prescribe boundary conditions on both the position of the boundary of the sheet and its orientation. We use Eq. \eqref{eq:mat} to replace the derivatives of the unit vectors by expressions containing only $\kappa$ and the unit vectors themselves. One can further eliminate the Lagrange multipliers from the vectorial Euler Lagrange equations to obtain a second order ordinary differential equation for the normal curvature:
\beq
\kappa''+\tfrac{1}{2}\kappa^3+C \kappa=0.
\label{eq:Elastica}
\eeq
As noted in \cite{CM05,MBAG08} the equation above is the same as the equation for equilibrium of the classical {\em Elastica}, and its solution may be given explicitly in terms of elliptic functions.
We choose $s=0$ at the center of the free standing portion of the sheet and obtain
\beq
\kappa(s)= d \cdot JacobiCN(\tfrac{d}{2c}s,c^2),
\label{eq:kgeneral}
\eeq
where, $JacobiCN(x,c^2)$ is the cosine of the Jacobi amplitude of the variable $x$ and the elliptic modulus $c$, and $-s_0\le s \le s_0$ \cite{Weisstein:fk}. The constants $d, c$ and the takeoff point $s_{0}$ are yet to be determined by the relevant boundary conditions; continuity of $\hatr, \hatt$ and of $\kappa$ across the takeoff line $s=\pm s_0$. These correspond to a continuous solution with no folds in which the normal force exserted by the supporting plane may be singular but not negative \cite{CM05,MBAG08}.

As the portion of the thin sheet that is supported on the plane has a vanishing curvature, we obtain the boundary conditions $\kappa(\pm s_0)=0$. This condition quantizes the values the product $d s_0/2 c$ can assume to the nodes of the $JacobiCN$ function. To be consistent with the observed form of the solution (concave center flanked by two convex regions, see Figure \ref{fig:cones}(f)) we set $d=s_{0}^{-1}6 c K(c^2)$, where $K(c^2)$ is the elliptic quarter period. Continuity of the tangent vector and the Gaussian charge $\d$ determine the two remaining constants $s_0$ and $c$. This is usually done via a shooting method \cite{CM05}. Practically, one may seek such solutions in the following fashion:
We choose a  co-ordinate system such that at $s=0$ $\hatr(0)=(1,0,0)$ and $\hatt(0)=(0,1,0)$.  Given these initial conditions, we have a two-dimensional family of solution for the set of equations \eqref{eq:mat}, corresponding to different choices for the values of $s_0$ and $c$. We denote by $2\theta_0$ the angle between $\hatr(s_0)$ and $\hatr(-s_0)$, as measured on the supporting plane. This angle can be related to the Lagrangian angle, $2 s_0$, measured along the sheet through the Gaussian  charge, $\d$, via the arc length condition $\theta_0=s_0-\d/2$. We expect the arc-length condition to single out one curve in the $(s_0,c)$ plane. An additional condition requires the vectors $\hatn(s_0)$ and $\hatn(-s_0)$ to coincide, since $\hatn(\pm s_0)$ is the normal to the supporting plane. The angle $\theta_n$  between these normal vectors is plotted in Figure \ref{fig:takeoff-normals}. According to this numerical  solution,  whenever $\hatn(s_0)=\hatn(-s_0)$ then $\hatr(s_0)=-\hatr(-s_0)$. 

This result implies in particular that the angle between takeoff lines is independent of the defect magnitude $\d$, This result can be understood in simple terms by considering the torque balance about the vertex of a plane supported cone. 
For a classical d-cone (with a vertical axis), the inward pressure from the contacting support exerts a torque about the apex.  The torque vector is in the horizontal plane and is normal to the plane of symmetry.  The necessary balancing torque comes from the singular forces at the take-off lines.  These torques are also horizontal and are normal to the two takeoff lines.  These two takeoff lines must have a nonzero component in the symmetry plane in order to produce a net torque.

The zero total torque condition may be used, together with vanishing of the total force on the system, to predict the angle between the takeoff lines for standard d-cones \cite{CM05}. In the present case, however, the portion of the elastic disk supported on the plane, does not change orientation and cannot exert any normal pressure on the supporting plane. Therefore, the singular force density at the takeoff lines cannot result in any torque, necessitating a takeoff angle of $\pi$. This also explains similar results obtained for other systems exhibiting planar self contact, which can be observed for example in Fig. 1 in ref \cite{SWBAM10}.

\section{Verifications}\label{Verifications}
To verify the prediction of the takeoff angle we minimize numerically a variant of the elastic energy functional of Equation \eqref{eq:lagrangian}. We eliminate the first Lagrange multiplier  by employing the polar angles of $\hatr$ as the unknown, and replace the second and third Lagrange multipliers by an in-plane elastic stretching term. To account for the impenetrability of the supporting plane we added a  short range repelling interaction decaying exponentially with the distance from the plane. 
%
Changing the integrated Gaussian charge by over an order of magnitude results in no discernible change in the takeoff angle, which remains constant near $180^\circ$, as observed in Figure \ref{fig:glued}(a).

The takeoff angle is also visualized experimentally using total internal reflection imaging allowing direct observation of the takeoff line. While due to extreme sensitivity of the system to its boundary conditions the system displays a deviation of about $10\%$ from the expected value of $180^\circ$, it shows no systematic variation with the opening angle $\delta$.

As noted above, the takeoff angle ceases to remain fixed when the support becomes conical; in the case of d-cones the (real space) takeoff angle shows quadratic variation with the parameter $\epsilon$ measuring the degree of confinement \cite{Wit07,CM05}. We now develop a framework which interpolates between the different kinds of conical defects described above: the unconstrained disclination, the d-cone and the plane supported e-cone

\section{Supported conical defects}\label{Supported}
\begin{figure}[ht]
\begin{center}
\includegraphics[width=8.6cm]{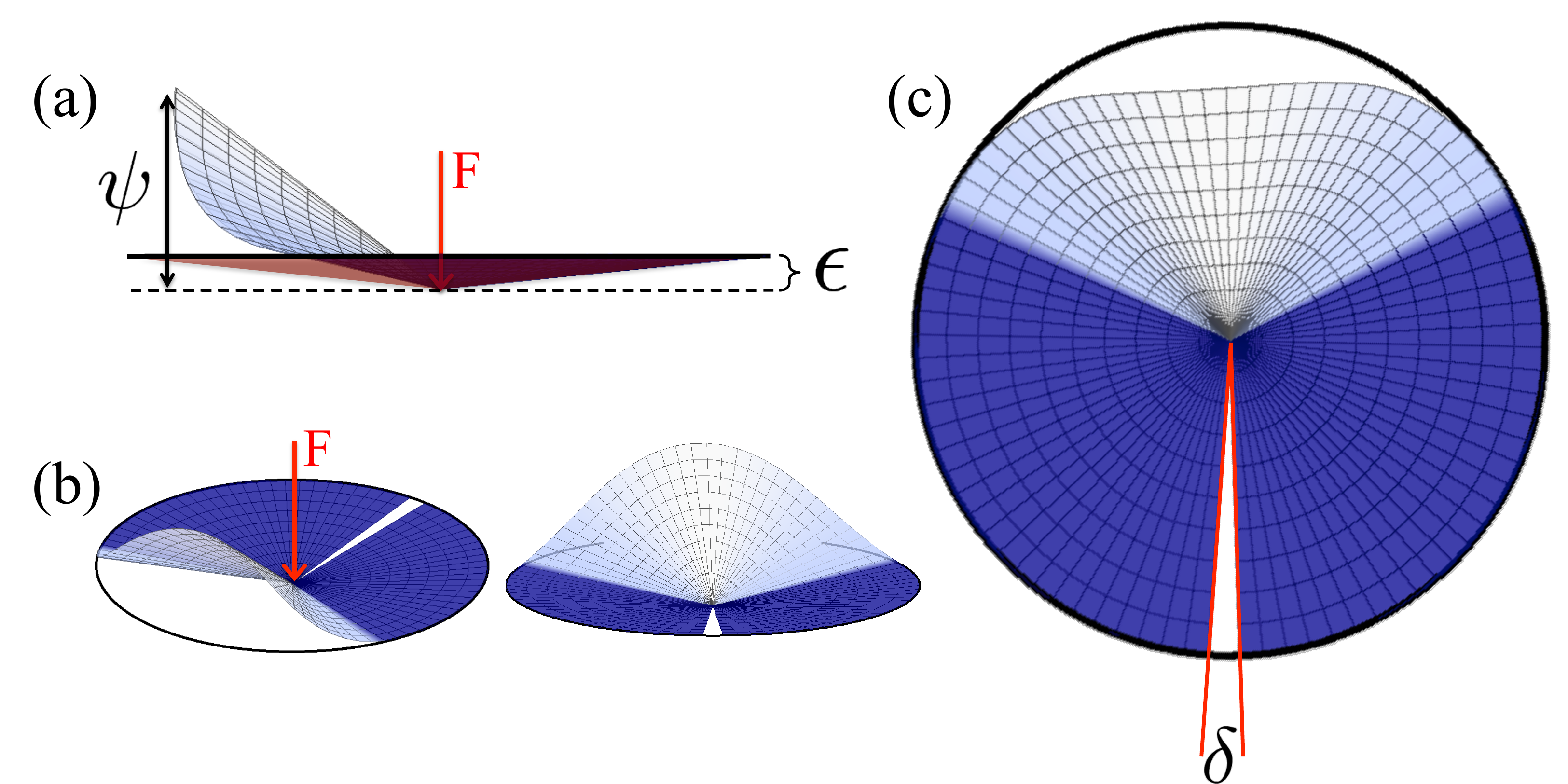}
\end{center}
\caption{\textit{The geometry of a supported conical defect:} (a) Side view; a notched disk is pushed into an impenetrable cone, shaded (red) in plot, by a force $F$ pushing down at its center point.  (b) Perspective view; colors correspond to vertical displacement of the edge. The dark (blue) region is in contact with the supporting cone. (c) Top view showing the opening angle of the notch, $\d$.  The (black) solid circle shows the perimeter of the supporting cone. }
\label{fig:generalSCD}
\end{figure}

We consider a general wedge disclination of arbitrary Gaussian charge 
$\d$ confined by an impenetrable frictionless cone of depth $\epsilon$ and base radius $1$, as depicted in Fig. \ref{fig:generalSCD}. The vertex of the conical defect and that of the confining cone are constrained to coincide.
The curvature $\kappa(s)$ in the free-standing buckled portion of the sheet is again given by the solution \eqref{eq:kgeneral}. However, all three constants $d,c$ and $s_0$ are now to be determined numerically. In order to gain insight into the the system we will next consider the linearized setting within the small slope approximation, which will allow explicit determination of all the relevant parameters in the solution.


\subsection{Small slope approximation}
In the small slope approximation the profile of the conical defect can be fully described using the height function $\psi= r\cdot \hat{z}$, see Fig \ref{fig:generalSCD}.  To leading order, the curvature in this case is given by, $\kappa=\psi+\psi''$ \cite{CM98}. Omitting the $\kappa^3$ term from Eq. \ref{eq:Elastica} leads to the linear equation 
\beq
\psi''''+(a^2+1)\psi''+a^2\psi=0.
\label{eq:LINode}
\eeq  
Here $a$ is a measure of the hoop compression. The height function $\psi(s)$ must satisfy the boundary conditions $\psi(\pm s_0)=\ep$ (continuity of the sheet) and $\psi'(\pm s_0)=0$ (no kinks, finite bending energy). 
Since the equation for $\psi(s)$ is linear and autonomous, the general solution can be expressed as a superposition of trigonometric functions:
\beq
\psi=\frac{\ep}{\cot(a s_0)-a \cot(s_0)}\Bigl( \frac{\cos(a s)}{\sin(as_0)}-a \frac{\cos(s)}{\sin(s_0)}\Bigr).
\label{eq:LINsolution}
\eeq
Here the two parameters appearing above; $a$, and the takeoff point $s_0$ are determined by the total length of the curve (which corresponds to determining $\d$), and the torque applied at the takeoff point, $\tau_0$. The form of the above solution is unique provided that $a^2\ne 1$ and $s_0\ne\pi/2$.
Noting that the curvature is continuous at the takeoff point $s_0$, we infer $\kappa(s_0) = \psi(s_0) +\psi''(s_0) = \epsilon$.   Together with the length restriction we obtain for $s_0$ and $a$:\footnote{These conditions yield infinitely many solutions for $a$. To be consistent with experimental results, as well as to minimize the total bending energy, we chose the first non-trivial solution to the equation; {\it i.e.}, the minimal value of $1<a$ which solves the equation.}
\beq
\begin{aligned}
&\frac{\d}{\ep^2}=s_0-\pi+\frac{s_0+\tan(s_0)\bigl(1-2 a^2+a^2 s_0 \tan(s_0)\bigr)}{2(a^2-1)},\\
& \tan(a s_0)/a s_0=\tan(s_0)/s_0.
\label{eq:s0&d}
\end{aligned}
\eeq
Note that the internal and external parameters $\delta$ and $\epsilon$ enter only through the ratio $\d/\ep^2$. This implies that the rescaled vertical displacement $\psi/\ep$ does not depend on the internal and external constraints
separately but only through the combination $\d/\ep^2$. Thus in the general case within the small slope approximation confinement and Gaussian charge are {\em interchangeable}. However, in two limiting cases, reviewed next, this is not true.

\begin{figure}[t]
\begin{center}
\includegraphics[width=8.6cm]{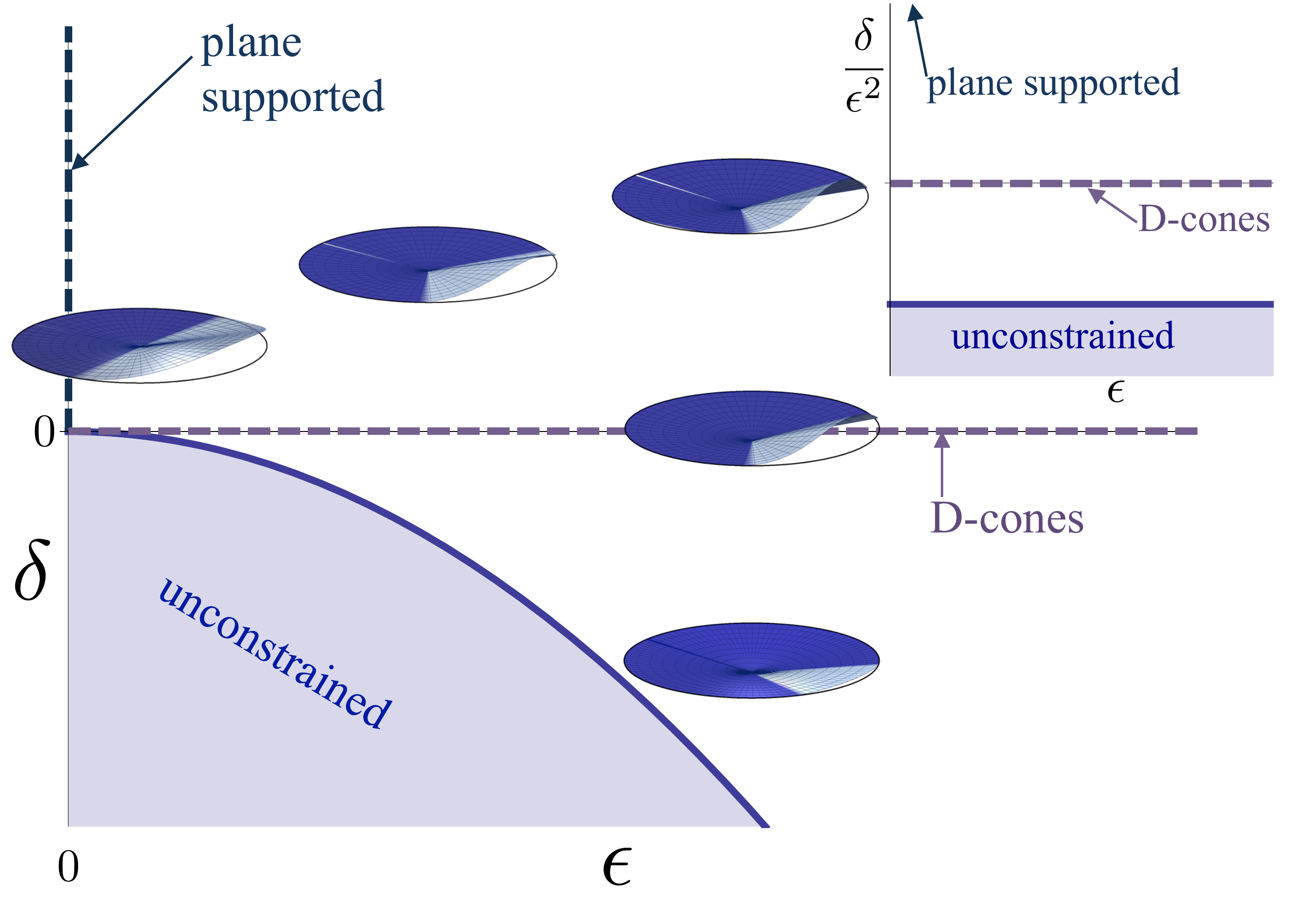}
\end{center}
\caption{\textit{Phase diagram showing how $\delta$ and $\epsilon$ affect the buckled shape}: Configurations in the shaded region are not constrained. The solid blue line marks the marginally constrained configurations satisfying $\d/\epsilon^2=-\pi$ and displaying a vanishing takeoff angle. The horizontal dashed line where  $\d=0$ corresponds to the well studied d-cones displaying a takeoff angle of about $139^\circ$, whereas the vertical dashed line corresponds to plane supported conical defects displaying a takeoff angle of $180^\circ$. Plotted in the $\d,\epsilon$ plane these three lines, corresponding to distinct behaviors, all meet at the origin. The inset shows the same phase diagram with $\delta$ replaced by $\delta/\epsilon^2$ on the vertical axis. }
\label{fig:phasediag}
\end{figure}

Given a finite (but sufficiently small) value for $\d$ we may obtain any value, arbitrarily large or small, for the parameter $\frac{\d}{\ep^2}$, by appropriately adjusting the value of $\ep$. Similarly, given a sufficiently small value for $\ep$ we can choose $\d$ to arrive at a desired valued for the combined parameter. The remaining possibilities are $\d=0$ and $\ep=0$. The case of vanishing Gaussian charge, $\d=0$, gives the familiar d-cone. In this case equation \eqref{eq:s0&d} does not depend on $\epsilon$ and for example the value $s_0=const\approx 139^\circ$ (measured on the deformed sheet). The case where $\ep=0$ corresponds to the plane supported conical defect. In this case equation \eqref{eq:s0&d} becomes singular. As we expect to approach this solution continuously as $\ep\to 0$ we are led to conclude that either $a^2=1$ or that $s_0=\pi/2$. The former corresponds to no hoop stress in the sheet and leads only to trivial solutions. Setting $s_0=\pi/2$ gives $a=3$ and leads to the explicit linearized solution
\footnote{
We note that to eliminate the torque on the supported disc, the Eulerian takeoff angle (measured on the supporting plane) must be $\pi$. Since in the present solution for the takeoff line the Lagrangian coordinate $s_0$ 
(measuring the amount of material n the free standing portion) is kept constant, the vanishing torque condition is satisfied only to leading order.
}
\beq
\psi=\sqrt{\frac{8\d}{\pi}}\cos(s)^3.
\label{eq:LINsol2}
\eeq
Since the amplitude $\psi$ is sublinear in $\delta$, $\psi$ can be substantial even with small $\delta$, as illustrated in Fig. \ref{fig:generalSCD}. 
These regimes are summarized in Fig \ref{fig:phasediag}.

\subsection{Analysis of the Takeoff angle}
\label{takeoff}

\begin{figure*}[ht]
\begin{center}
\includegraphics[width=17.2cm]{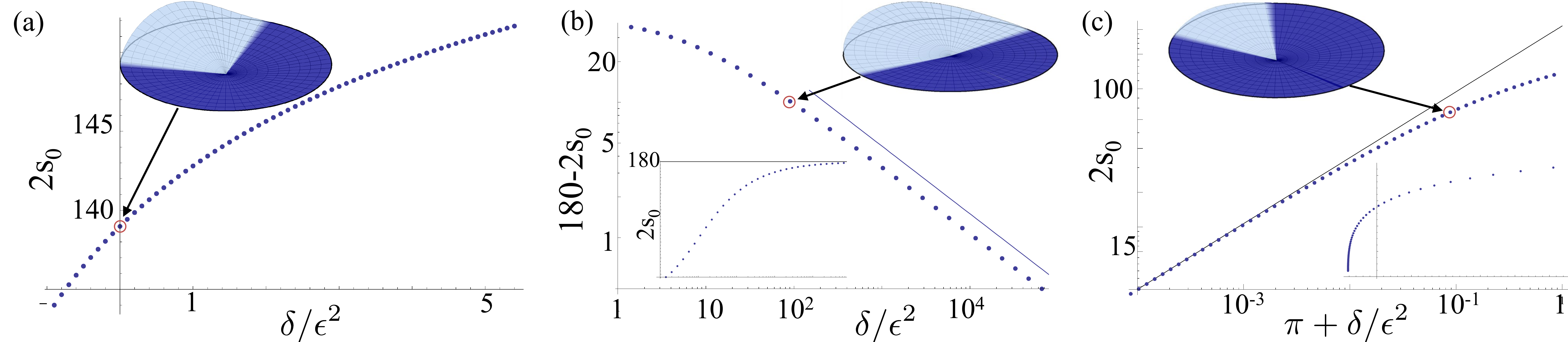}
\end{center}
\caption{\textit{The takeoff angle:} Plots show behavior of takeoff angle $s_0$ with parameter $\delta/\epsilon^2$ in different ranges.  Curves were calculated by solving \eqref{eq:s0&d} numerically. The inset disks were obtained by numerically minimizing the quadratic energy functional ({\it e.g.} Eq. 5 of Ref. \cite{CM98}) that generates Eq. \eqref{eq:LINode}, replacing the Lagrange multipliers by appropriate (smooth and steep) potentials, verifying that the details of the potential do not affect the resulting equilibrium.
(a) Moderate $\delta/\epsilon^2$. For small values of $\d/\epsilon^2$ the classic d-cone takeoff angle $\sim 139^\circ$ is slightly perturbed. Increasing (decreasing) the value of $\d/\ep^2$, increases (decreases) the value of $s_0$.  (b) Large $\delta/\epsilon$.  (Inset shows global behavior.)  For large values, $1\ll \d/\ep^2$ an upper limit of $2s_0=180^\circ$ is found. The approach to this limiting value follows  $180-2 s_0\propto \ep/ \sqrt{\d}$. (c) Negative $\delta$. When $\delta/\epsilon^2 < -180^\circ$  the sheet is not constrained. As this limiting value is approached, $ s_0 \propto (\pi-\d/\ep^2)^{1/3}$. Thus delicate tuning is needed to create small takeoff angles.  Inset shows global behavior.
}
\label{fig:takeoff}
\end{figure*}

The takeoff angle, $2s_0$, measures the amount of material in the free standing portion of the disk. Fig. \ref{fig:takeoff} shows its dependence on $\delta/\epsilon^2$. A takeoff angle of $2 s_0$ measured on the disk corresponds to a complementary angle (length at radius unity) of $2\pi+\d-2 s_0$ resting on the supporting cone. Therefore the real space opening between the takeoff lines, to leading order in $\d$ and $\ep$, reads
\[
2 \theta_0=2 s_0 -\d- \ep^2(\pi-s_0).
\]
We consider a small but non-vanishing value of $\ep$, and a small $\d$ such that the small slope approximation made above remains valid, while obtaining a variety of values for $\d/\ep^2 $.
Considering $\d/\ep^2\ll 1$ we perturb the takeoff angle solution obtained for the d-cone ($\d=0$) only slightly. As seen in figure \ref{fig:takeoff}(a) increasing the value of $\d$ (adding material) increases $s_0$, whereas decreasing the value of $\d$ (to assume negative values) lowers $s_0$.
One may also consider cases in which $1\ll \d/\ep^2$. As this parameter tends to infinity the value of the takeoff angle approaches $\pi$, the result obtained for plane supported conical defects ($\ep=0$) above. The variation about this limiting value scales with the square root of the parameter : $\pi - 2 s_0\propto \ep/ \sqrt{\d}$. 

Considering negative values of $\d$, the takeoff angle diminishes as the parameter $\d/\ep^2$ becomes larger in absolute value until it reaches the lower bound $\d/\ep^2=-\pi$ beyond which the cone is unconstrained as its natural radius (with a circular cross section) is smaller than that of the confining cone. The takeoff angle reduces to zero at $\d/\ep^2=-\pi$ and scales as $ s_0 \propto (\pi-\d/\ep^2)^{1/3}$.

In particular the above implies that the simultaneous limit $\epsilon\to 0$ and $\d\to 0$ depends on the order of limits, or more accurately on the value of $\d/\epsilon^2$. As seen in Figure \ref{fig:phasediag} the three thick lines correspond to marginally confined thin sheets ($\d/\epsilon^2=-\pi$),  d-cones ($\delta=0$), and plane supported conical defects, ($\epsilon=0$). These in turn correspond to takeoff angles of $139^\circ$ and $180^\circ$, respectively.    

\section{Discussion}\label{Discussion}

The derivations above suggest that thin sheets with negative disclinations respond distinctively when confined by planar boundaries.  Such sheets are a limiting case of generic disclinations confined by a cone.  In all these cases, stresses owing to the confinement cause the sheet to buckle away from the confining surface.  We have shown that the generic sheet responds smoothly to changes in the confining shape ({\it i.e.}, the cone angle $\epsilon$) or the disclination charge ({\it i.e.}, the wedge angle $\delta$).  Further, these two changes are equivalent when perturbations are small and the sheet is nearly flat. Only the combination $\delta/\epsilon^2$ affects the shape.  When the confining surface becomes planar, this control parameter becomes infinite.  In this limit the takeoff angle approaches $\pi$, in agreement with the specific analysis of Sec. \ref{Plane}.  However, the control parameter cannot predict the buckling amplitude in this planar limit; a more delicate analysis is required to obtain this amplitude.  
					
					Our predictions are based on an idealization of the surface in which we neglect any tangential strain in the sheet and thus take the sheet to be strictly isometric. This assumption greatly simplifies our analysis: it allows us to describe the sheet as a generalized cone \cite{CM05,MBAG08}, whose only degree of freedom is the reduced curvature $\kappa(s)$.  It is natural to anticipate that any real sheet approaches the conical shape as its thickness goes to zero.  However, the nature of this convergence has not been fully established, despite serious efforts \cite{Gemmer:2012kx,Brandman:2013vn}.  Poor convergence could limit the applicability of our predictions to real sheets.
					
					This planar case is unique because only here is the width of the buckled region independent of its amplitude.  For this reason there is particular interest in observing this independence experimentally.  We were able to observe the contact line clearly in experiments. We found that the takeoff angle is insensitive to the amplitude or the opening angle.  Still, the angle itself remained different from expected value of $\pi$.   In our experiment the takeoff lines form an angle of about $160^\circ$ rather than $180^\circ$. The takeoff angle is extremely sensitive to variations in the setup. In particular friction with the supporting surface, relative shearing of the clamped edges of the disk as well as finite core effects all give rise to a smaller takeoff angle. While in specific systems takeoff angles above $175^\circ$ were observed, the image depicted in Fig. \ref{fig:glued} is a typical representative of the general result. This discrepancy suggests that our mathematical idealizations miss a needed ingredient in the case of planar confinement.  
					
					The real sheets were not infinitely thin, and thus they deform by stretching as well as bending.  The experimental behavior of the sheet gave hints of such stretching.  In the experiment the vertex is forced onto the surface with a stylus.  Without sufficient force the vertex rises off the plane.  The force is larger than would be expected from simply bending the sheet with the required strong local curvature.  This leads to the suspicion that stretching occurs near the vertex, thus distorting the conical shape there.  While our numerical simulations allowed for nonzero stretching (and deformability of the plane), they did not allow deviation from the conical shape.  In order to control this form of buckling and remain within the regime of negligible stretching, some way to reduce the strong forcing needed in the experiment seems necessary.  For example, one might allow the confining plane to approach the vertex but not to touch it.  

					The buckling phenomena shown here greatly generalize the known cases of confinement-induced deformation.  Our analysis may thus may be of value in designing new kinds of actuators.  For example, the plane-confined defect converts motion in a plane (via increasing $\delta$) to amplified motion and forcing normal to that plane (via the buckled region).  As we have seen, opening the wedge converts a flat sheet to a three-dimensional sheet that subtends nonzero volume.  Since these actuators don't require explicit molding or bending material, they may prove especially useful for nanoscale devices made {\it eg} with graphene or other molecular sheets \cite{Blees:2014ys}.  The simple cases shown here must have counterparts with more general confining surfaces.  Examples are developable surfaces more general than cones, or those lacking circular symmetry.  

					An interesting implication of the planar-confined e-cone studied above is that the confined shape does not require an entire plane.  We noted above that the flat region in contact with the sheet exerts no normal force on it up to the take-off line.  Thus this part of the plane may be removed without consequence.  It suffices to confine the e-cone on a narrow wedge and on a perpendicular line in the plane of the wedge (the takeoff line).  

					Finally, the generalizations of the confined defects worked out here provide potential leverage to attack an outstanding puzzle regarding the d-cone \cite{Wit07}.  D-cones are shown to have a core region governed by the thickness of the sheet, and thence by tangential stress.  However, the nature of this stress and how it controls the size of the core region remain a mystery.  The nature of the core region for marginally-confined cones with $s_0$ approaching zero would surely give insight into how the core arises.

\section{Conclusion}\label{Conclusion}
					This study extends our knowledge of the relationship between external and internal constraints on a thin sheet.  We find that under mild conditions the two forms of constraint have equivalent effects.  These findings add to our flexibility in creating new forms of self-shaping materials.  They also raise the question of how far this equivalence extends to stronger perturbations.  Both subjects hold promise for future study.  

					\acknowledgements
The authors are grateful to William Irvine, Dustin Kleckner, Sidney Nagel, and Shmuel Rubinstein for experimental advice.  This work was supported by the Simons Foundation, the NSF Materials World Network program under Award Number DMR-0807012, and the University of Chicago MRSEC program of the NSF under Award Number DMR 0820054.

\newpage
\bibliography{SCD_lib2}

\end{document}